\documentclass{IEEEtran}
\usepackage{amsmath,amsfonts}
\usepackage{array}
\usepackage[caption=false,font=normalsize,labelfont=sf,textfont=sf]{subfig}
\usepackage{textcomp}
\usepackage{stfloats}
\usepackage{url}
\usepackage{verbatim}
\usepackage{graphicx}
\usepackage{cite}
\usepackage{hyperref}
\usepackage{booktabs}
\usepackage{makecell}
\usepackage{gensymb}
\usepackage{xcolor}
\usepackage{multirow}
\usepackage{enumitem}
\usepackage{graphicx} 
\usepackage{adjustbox}
\usepackage{algorithm}%
\usepackage{algorithmicx}
\usepackage{bm}
\usepackage{algpseudocode}

\numberwithin{algorithm}{section} 
 
\usepackage{threeparttable}
\begin{document}

\title{State Space and Self-Attention Collaborative Network with Feature Aggregation for DOA Estimation} 

\author{
    \IEEEauthorblockN{Qi You, Qinghua Huang, Yi-Cheng Lin}
    \thanks{Qi You and Qinghua Huang are with the School of Communication and Information Technology, Shanghai University, Shanghai 200444, China (e-mail: youq@shu.edu.cn; qinghua@shu.edu.cn). 
    Yi-Cheng Lin is with National Taiwan University, Taiwan 10617, China
 (e-mail: f12942075@ntu.edu.tw).}
}


\maketitle

\begin{abstract}
Accurate direction-of-arrival (DOA) estimation for sound sources is challenging due to the continuous changes in acoustic characteristics across time and frequency. In such scenarios, accurate localization relies on the ability to aggregate relevant features and model temporal dependencies effectively. In time series modeling, achieving a balance between model performance and computational efficiency remains a significant challenge. To address this, we propose FA-Stateformer, a state space and self-attention collaborative network with feature aggregation. The proposed network first employs a feature aggregation module to enhance informative features across both temporal and spectral dimensions. This is followed by a lightweight Conformer architecture inspired by the squeeze-and-excitation mechanism, where the feedforward layers are compressed to reduce redundancy and parameter overhead. Additionally, a temporal shift mechanism is incorporated to expand the receptive field of convolutional layers while maintaining a compact kernel size. To further enhance sequence modeling capabilities, a bidirectional Mamba module is introduced, enabling efficient state-space-based representation of temporal dependencies in both forward and backward directions. The remaining self-attention layers are combined with the Mamba blocks, forming a collaborative modeling framework that achieves a balance between representation capacity and computational efficiency. Extensive experiments demonstrate that FA-Stateformer achieves superior performance and efficiency compared to conventional architectures.
\end{abstract}


\section{Introduction}
\IEEEPARstart{S}{ound} source localization (SSL) refers to the task of estimating the spatial positions of acoustic sources by processing multi-channel acoustic signals captured by microphone arrays. In practice, SSL is often formulated as direction-of-arrival (DOA) estimation, which aims to determine the incoming angles of signal sources.
Accurate DOA estimation plays a crucial role in a wide range of applications, such as audio surveillance in industrial environments \cite{changFeatureExtractedDOA2018}, underwater acoustic communications \cite{xuTwoDimensionalDOAEstimation2025}, and autonomous driving \cite{houRobustDOATracking2025}.

Over the past decades, a variety of algorithmic frameworks have been developed to tackle this problem from different perspectives. Among them, subspace-based algorithms such as multiple signal classification (MUSIC) \cite{wangImprovedMultipleSignal2020} and the estimation of signal parameters via rotational invariance techniques (ESPRIT) \cite{fang-minghanESPRITlikeAlgorithmCoherent2005} are well-known for their ability to provide high-resolution results in ideal scenarios. Beamforming strategies, including steered response power with phase transform (SRP-PHAT) \cite{cobosModifiedSRPPHATFunctional2011} and minimum variance distortionless response (MVDR) \cite{sunJointDOATDOA2011}, achieve localization by evaluating spatial response patterns, showing strong performance in environments with limited reverberation. Another widely used class of techniques is based on time difference of arrival (TDOA). Methods such as the generalized cross-correlation with phase transform (GCC-PHAT) \cite{kwonAnalysisGCCPHATTechnique2010} infer direction from estimated inter-sensor delays. 
While these traditional methods work well in controlled environments, their accuracy drops in the presence of reverberation and noise. In addition, they perform poorly when there are multiple sources or when the sources are moving.

In recent years, deep learning has transformed the field of SSL. Moving beyond traditional signal processing frameworks that depend on explicit physical assumptions, data-driven approaches learn spatial and temporal structures directly from multichannel observations, demonstrating remarkable adaptability in complex and reverberant environments. 
A variety of architectures have been explored, spanning from convolutional and recurrent networks to more recent Transformer and Mamba architectures \cite{diaz-guerraRobustSoundSource2021,yangStackedSelfattentionNetwork2022,yinMIMODoAnetMultichannelInput2022a,yangSRPDNNLearningDirectPath2022a,wangFourStageDataAugmentation2023,tangSoundSourceLocalization2023,zhuIFANIcosahedralFeature2024,wangIPDnetUniversalDirectPath2024a,songgongMultiSpeakerLocalizationCircular2024a,dwivediSparseBayesianIntegrated2025,baekDNNBasedGeometryInvariantDOA2025,xiaoTFMambaTimeFrequencyNetwork2025,pi2026modal}. A more detailed discussion is presented in Section~\ref{sec:Related Works}.
Broadly speaking, current deep learning-based methods can be categorized into non-end-to-end and end-to-end methods. Non-end-to-end methods do not perform direct localization. Instead, they aim to assist traditional DOA estimation frameworks by learning or utilizing intermediate representations. 
In contrast, end-to-end frameworks directly learn a mapping from multichannel observations to source positions, integrating feature learning and localization into a unified model. This paradigm simplifies the overall pipeline and demonstrates strong generalization across different acoustic conditions. 

However, most existing systems remain constrained by assumptions of static or single-source scenarios. Handling multiple or dynamically moving sources introduces additional complexity, as the spatial and temporal dependencies in such scenes challenge both network design and training strategies. Although some recent studies have attempted to extend deep learning models to track moving sources \cite{diaz-guerraRobustSoundSource2021,zhuIFANIcosahedralFeature2024,yangSRPDNNLearningDirectPath2022a,wangFourStageDataAugmentation2023,wangIPDnetUniversalDirectPath2024a,xiaoTFMambaTimeFrequencyNetwork2025,pi2026modal} these methods introduce new challenges that limit their practicality:

\begin{enumerate}[label=\arabic*)]

    \item Conventional sequence modeling methods, including recurrent and Conformer-based architectures, face challenges in balancing temporal modeling capacity and computational efficiency. Particularly in real-time or resource-constrained settings, these methods often incur significant overhead, making them less practical for deployment.

    \item Mamba-based models are well-suited for capturing long-range dependencies with high efficiency, but they tend to be less sensitive to subtle local changes. This shortcoming can affect DOA estimation accuracy in dynamic conditions, especially when the source direction undergoes rapid short-term variations.
\end{enumerate}

\begin{table*}[t]
\centering
\resizebox{\linewidth}{!}{%
\begin{threeparttable}
\caption{Summary of Deep Learning-Based Method Types}
\label{tab:localization_methods}
\begin{tabular}{>{\centering\arraybackslash}c
                >{\centering\arraybackslash}c
                >{\centering\arraybackslash}c
                >{\centering\arraybackslash}c
                >{\centering\arraybackslash}c
                >{\centering\arraybackslash}c
                >{\centering\arraybackslash}c}
\hline\hline
Reference & Year & Model & Type & Input Features & Output & NoS\tnote{1} \\
\midrule
\cite{diaz-guerraRobustSoundSource2021} & 2021 & CNN & Regression & SRP-PHAT & \(x,y,z\) & 1 \\
\cite{yangStackedSelfattentionNetwork2022} & 2022 & Attention & Regression & STFT coefficients & \(x,y\) & 1--2\\
\cite{yinMIMODoAnetMultichannelInput2022a} & 2022 & GRU & Regression & Magnitude spectrogram, IPD & SPS & 3--4\\
\cite{yangSRPDNNLearningDirectPath2022a} & 2022 & CRNN & Regression & Phase and magnitude spectrograms & DP-IPD & 1--2\\
\cite{wangFourStageDataAugmentation2023} & 2023 & ResNet-Conformer & Regression & log-Mel spectrogram, GCC-PHAT & \(x,y,z\) & 1--3\\
\cite{tangSoundSourceLocalization2023} & 2023 & CNN & Regression & Time-domain sampling point & \(\theta, \phi\) & 1--2\\
\cite{zhuIFANIcosahedralFeature2024} & 2024 & Icosahedral CNN & Regression & \makecell{SRP-PHAT, SRP-LMS} & \(x,y,z\) & 1\\
\cite{wangIPDnetUniversalDirectPath2024a} & 2024 & LSTM & Regression & STFT coefficients & DP-IPD & 1--2\\
\cite{songgongMultiSpeakerLocalizationCircular2024a} & 2024 & CNN & Classification & Circular harmonic feature & \(\theta\) & 2\\
\cite{dwivediSparseBayesianIntegrated2025} & 2025 & CNN, SBL & Classification & Spherical harmonic feature & \(x,y,z\) & 1\\
\cite{baekDNNBasedGeometryInvariantDOA2025} & 2025 & MHSA, GRU & Classification & STFT and spherical coordinates & SPS & 1\\
\cite{xiaoTFMambaTimeFrequencyNetwork2025} & 2025 & Mamba & Regression & STFT coefficients & SPS & 1\\
\cite{pi2026modal} & 2026 & CRNN, LSTM & Classification & STFT coefficients & \(\theta\) & 1\\
\midrule
Proposed & - & Conformer, Mamba & Regression & Phase and magnitude spectrograms & \(x,y,z\) & 1--2\\
\bottomrule
\end{tabular}
\begin{tablenotes}
\footnotesize
\item[1] NoS: considered number of sources.
\end{tablenotes}
\end{threeparttable}
} 
\end{table*}

To deal with the above problems, we propose a feature aggregation enhanced state space and self-attention collaborative network (FA-Stateformer), specifically designed for DOA estimation. The key contributions of this work are as follows.
\begin{enumerate}[label=\arabic*)]
    \item To balance modeling capability and computational efficiency, we designed a sequence modeling framework that collaborates Mamba-based bidirectional state space modeling with self-attention layers to jointly capture global dependencies and local dynamic features.
    
    \item A lightweight Conformer backbone is designed by introducing a feedforward compression mechanism along with a temporal shift operation in convolutional layers, enabling the model to capture long-range dependencies more effectively while maintaining minimal computational overhead.
    
    \item Extensive experiments on both simulated and real-world datasets demonstrate the superiority of the proposed FA-Stateformer. The model achieves higher accuracy and efficiency than existing methods. Ablation studies further validate the contribution of each individual module.
    
\end{enumerate}

The rest of this paper is organized as follows. In Section~\ref{sec:Related Works}, recent deep learning-based DOA estimation methods using microphone arrays are reviewed. The proposed design is detailed in Section~\ref{sec:Methodology}.  Experimental results are presented in Section~\ref{sec:Experiments and Discussions}, including the experimental setup and analysis.  Finally, the paper concludes in Section~\ref{sec:Conclusion} with directions for future work.

\section{Related Works}
\label{sec:Related Works}

\subsection{Deep Learning-Based DOA Estimation for Sound Sources Localization}

Deep learning has significantly advanced DOA estimation, especially under complex conditions involving moving sound sources. Representative approaches proposed in recent years are summarized in Table~\ref{tab:localization_methods}. Existing methods typically use either raw multi-channel audio or features obtained through classical signal processing. Common representations include Fourier-based time-frequency spectra~\cite{yangStackedSelfattentionNetwork2022,yangSRPDNNLearningDirectPath2022a,wangIPDnetUniversalDirectPath2024a,songgongMultiSpeakerLocalizationCircular2024a,pi2026modal,baekDNNBasedGeometryInvariantDOA2025,dwivediSparseBayesianIntegrated2025,xiaoTFMambaTimeFrequencyNetwork2025}, inter-channel phase or amplitude differences~\cite{yinMIMODoAnetMultichannelInput2022a}, sound intensity vectors~\cite{wangFourStageDataAugmentation2023}, and cross-correlation functions~\cite{diaz-guerraRobustSoundSource2021,zhuIFANIcosahedralFeature2024,wangFourStageDataAugmentation2023}. Recent studies have directly used time domain sampling points as network input for sound source localization without relying on any basic signal processing algorithms~\cite{tangSoundSourceLocalization2023}.
The outputs of DOA estimation networks are usually formulated either as classification over discretized spatial grids~\cite{songgongMultiSpeakerLocalizationCircular2024a,tangSoundSourceLocalization2023,pi2026modal} or as continuous regression in Cartesian or spherical coordinates~\cite{diaz-guerraRobustSoundSource2021,yangStackedSelfattentionNetwork2022,wangFourStageDataAugmentation2023,zhuIFANIcosahedralFeature2024,dwivediSparseBayesianIntegrated2025}. While classification limits spatial resolution due to discretization, regression enables more precise localization by directly estimating continuous DOA values. In terms of network structures, researchers have investigated convolution-based models such as CNNs~\cite{diaz-guerraRobustSoundSource2021,tangSoundSourceLocalization2023,zhuIFANIcosahedralFeature2024,songgongMultiSpeakerLocalizationCircular2024a,dwivediSparseBayesianIntegrated2025}, ResNets~\cite{wangFourStageDataAugmentation2023}, and CRNNs~\cite{yangSRPDNNLearningDirectPath2022a,pi2026modal}, as well as recurrent models like LSTMs~\cite{wangIPDnetUniversalDirectPath2024a,pi2026modal,baekDNNBasedGeometryInvariantDOA2025} and attention-driven architectures~\cite{baekDNNBasedGeometryInvariantDOA2025}.
To further improve acoustic modeling, Wang et al.~\cite{wangFourStageDataAugmentation2023} incorporated the Conformer framework into ResNet and achieved the best results in the Challenge on Detection and Classification of Acoustic Scenes and Events (DCASE) 2022 challenge.
Recently, Xiao et al.~\cite{xiaoTFMambaTimeFrequencyNetwork2025} adopted a new architecture Mamba based on neural state space model (SSM) to realize single moving sound source localization and achieved performance better than the state-of-the-art models.
In summary, deep learning-based approaches have shown strong potential for accurate SSL. However, many methods remain less effective in complex acoustic environments involving multiple or moving sources. In addition, their training and inference often require considerable computational resources. For example, the IPDnet proposed by Wang et al.~ \cite{wangIPDnetUniversalDirectPath2024a} achieves high accuracy, but its high computational complexity limits its deployment on portable devices with limited hardware resources.

\subsection{State Space Models}
State Space Models (SSMs) have long been valued in sequence modeling for their ability to represent temporal dynamics through latent state transitions. Building on this foundation, the Mamba~\cite{guMambaLinearTimeSequence2024} architecture has emerged as a recent and notable advancement in state space modeling. TF-Mamba~\cite{xiaoXLSRMambaDualColumnBidirectional2025} is the first to apply Mamba to sound source localization, extending its modeling capability across both time and frequency domains to enhance spatial feature representation. oSpatial-Mamba~\cite{quanMultichannelLongTermStreaming2024} incorporated the state space framework into SpatialNet~\cite{quanSpatialNetExtensivelyLearning2024} for multi-channel speech enhancement, showing better performance in challenging acoustic conditions with both stationary and moving speakers. 
S-Mamba~\cite{wangMambaEffectiveTime2025} introduced a bidirectional structure to overcome the limitation of the standard Mamba.
Bi-Mamba+ ~\cite{liangBiMambaBidirectionalMamba2024} introduced a series-relation-aware decider to dynamically switch between channel-independent and channel-mixing strategies.

Despite these advances, most Mamba-based methods are still applied mainly to general time series tasks such as forecasting and classification, with relatively few studies targeting DOA estimation. In addition, existing work often highlights the selective recurrence mechanism of Mamba but pays less attention to how it could be combined with self-attention models such as the Transformer or Conformer. 
For example, ConMamba~\cite{houConMambaConvolutionAugmentedMamba2024} replaces the multi-head attention module in the Conformer with Mamba, but it does not further explore how a closer integration of the two frameworks could be designed.

\section{Methodology}
\label{sec:Methodology} 
\subsection{Problem Statement}
In reverberant indoor environments, the signal received by an array of \( C \) microphones is a mixture of multiple sound sources, each convolved with the corresponding room impulse response (RIR) and corrupted by noise. This process can be formulated as:
\begin{equation}
\label{eq:pf_ex1a}
y_c(n) = \sum_{s=1}^{S} x^{(s)}(n) * h^{(s)}_c(n) + v_c(n)\; ,
\end{equation}
where \(x^{(s)}(n)\) denotes the \(s\)-th source signal, \(h^{(s)}_c(n)\) is the RIR from the \(s\)-th source to the \(c\)-th microphone, \(v_c(n)\) is the additive noise, and \( * \) denotes convolution. 

To extract spatial information, the time-domain signals are converted into the time-frequency domain using an \(N\)-point short-time Fourier transform (STFT). The transformed signal at microphone \(c\) is written as \(Y_c(t,f)\), where \(t\) and \(f\) are the time frame and frequency bin indices. For DOA estimation, the phase spectrum is of particular importance, since inter-channel phase differences encode the spatial cues required for localization.
In practice, we compute the complex STFT coefficients and normalize them to obtain \(\overline{Y}_c(t,f)\). Both log-magnitude and phase spectra are computed, normalized, and arranged into a tensor of size \(C \times F \times T\), where \(C\) is the number of channels, \(F\) the number of frequency bins, and \(T\) the number of time frames.

\subsection{The Structure of FA-Stateformer}
\begin{figure}[!t]
\centering
\includegraphics[width=0.9\linewidth]{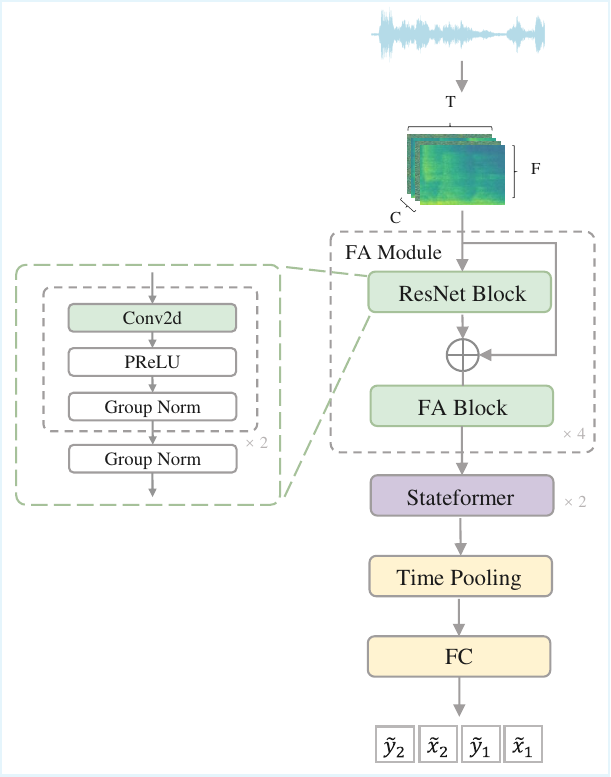}
\caption{The proposed FA-Stateformer for DOA estimation.}
\label{fig:FA-Stateformer}
\end{figure}
As shown in Fig.~\ref{fig:FA-Stateformer}, the feature aggregation (FA) module consists of two main components: a ResNet block for learning deep hierarchical representations and a FA block that refines the extracted features. The FA block enhances the quality of input representations by aggregating informative features across both time and frequency dimensions. This strategy has already been shown to be effective in our previous work~\cite{wangFA3NetFeatureAggregation2025}. The aggregated representation is then used as the input to the subsequent proposed Stateformer.

\subsubsection{Feature aggregation and enhancement}
\begin{figure}[!t]
    \centering
    \includegraphics[width=\linewidth]{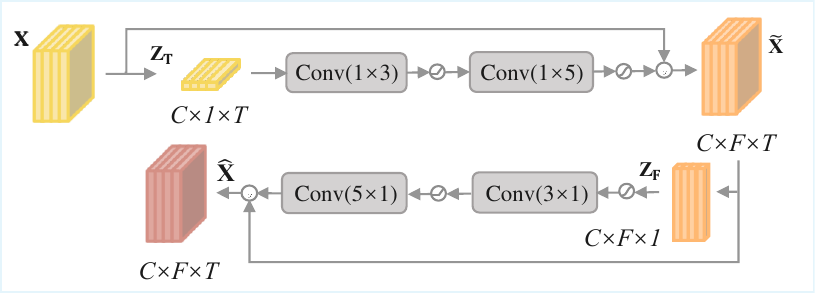}
    \caption{The proposed FA block. \(C\),\(T\) and \(F\) denote the dimension of channel, time and frequency respectively.}
    \label{fig:FAA}
\end{figure}

Spectrograms are the main input for DOA estimation, but they are quite different from images. In image tasks, spatial dimensions are continuous and nearby pixels are usually related. In audio spectrograms, however, the time and frequency axes have different physical meanings and do not always follow stable or consistent patterns. Because of this, directly applying two-dimensional modeling methods developed for images may create false links between unrelated time frames or frequency bands, which can lower the accuracy of localization.
To address this issue, we introduce a FA block that processes the time and frequency dimensions separately. As shown in Fig.~\ref{fig:FAA}, the block first compresses features along one dimension and then learns attention weights through lightweight strip-shaped convolutions, enabling the network to highlight informative structures while avoiding irrelevant correlations.  

For the temporal branch, features are averaged across frequency bins:
\begin{equation}
\mathbf{Z}_{\text{T}}(c,t) = \frac{1}{F}\sum_{f=1}^F \mathbf{X}(c,t,f)\; ,
\end{equation}
where \(\mathbf{Z}_{\text{T}}(c,t)\) denotes the aggregated representation at channel \(c\) and time step \(t\). Channel dependencies are then captured using successive convolutions with kernel sizes \(1 \times 5\) and \(1 \times 3\), followed by nonlinear activations:
\begin{equation}
\mathbf{W}_{\text{T}} = \sigma_s\big(f_{1 \times 3}\big(\sigma_r(f_{1 \times 5}(\mathbf{Z}_{\text{T}}))\big)\big),
\end{equation}
where \(\sigma_r\) and \(\sigma_s\) denote ReLU and Sigmoid functions, respectively, and \( f_{k \times k}(\cdot) \) represents a convolutional operation with a kernel size of \( k \times k \). The resulting temporal attention map is broadcast along the frequency dimension to reweight the original features:
\begin{equation}
\tilde{\mathbf{X}} = \mathbf{X} \odot \mathbf{W}_{\text{T}}\; ,
\end{equation}
where \( \odot \) denotes element-wise product.

A similar process is applied to the frequency branch. Features are first averaged across the temporal dimension:
\begin{equation}
\mathbf{Z}_{\text{F}}(c,f) =  \frac{1}{T}\sum_{t=1}^T \tilde{\mathbf{X}}(c,t,f),
\end{equation}
then passed through \(5 \times 1\) and \(3 \times 1\) strip convolutions with nonlinearities to obtain the frequency attention map:
\begin{equation}
\mathbf{W}_{\text{F}} = \sigma_s\big(f_{3 \times 1}(\sigma_r(f_{5 \times 1}(\mathbf{Z}_{\text{F}})))\big).
\end{equation}

Finally, both attention maps are applied sequentially to generate the refined representation:
\begin{equation}
\hat{\mathbf{X}} = \tilde{\mathbf{X}} \odot \mathbf{W}_{\text{F}}\; .
\end{equation}

This two-stage design ensures that the network adaptively emphasizes important cues along both time and frequency dimensions, while suppressing irrelevant patterns that could interfere with localization.

\subsubsection{Squeeze and excitation Conformer}

\begin{figure}[!t]
    \centering
    \includegraphics[width=\linewidth]{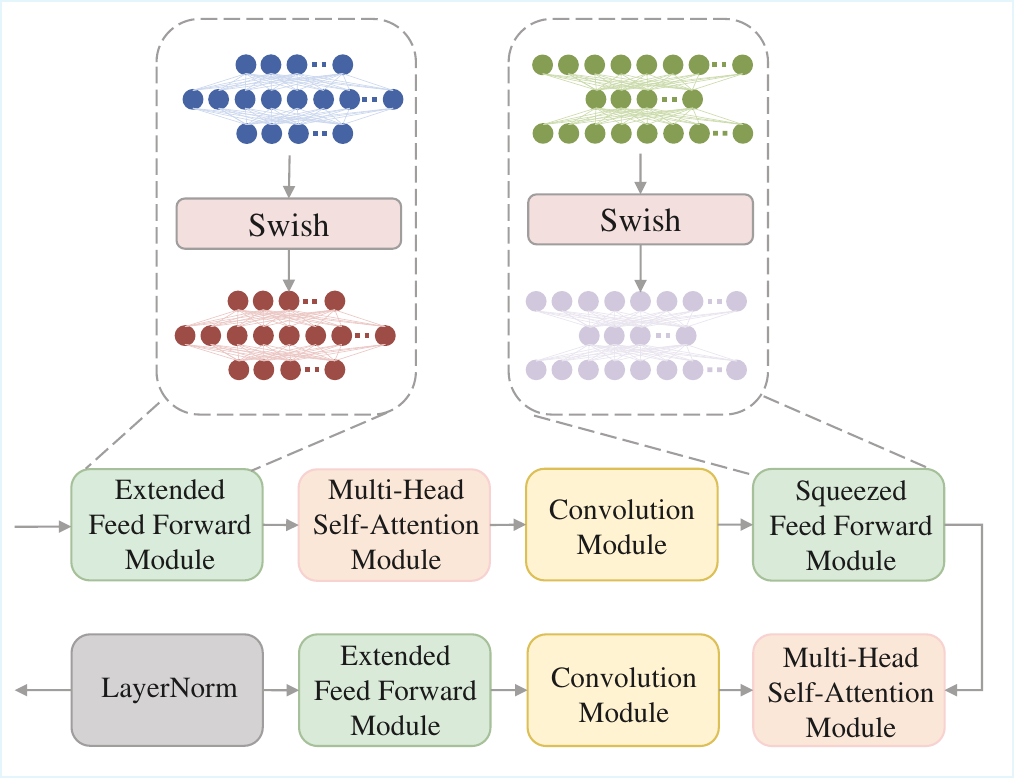}
    \caption{The framework of SEConformer.}
    \label{fig:conformer}
\end{figure}
The Conformer module~\cite{wangFourStageDataAugmentation2023} is an innovative combination of self-attention mechanism and convolution. The self-attention mechanism captures global dependencies, while the convolution learns local features in the audio sequence.
In conventional Conformer architectures, the convolutional module and the feed-forward network (FFN) are the two key components responsible for local dynamic modeling and feature transformation. Although this design achieves excellent performance in speech and audio-related tasks, its stacked structure introduces substantial computational redundancy, which limits inference efficiency and increases resource usage. To improve model adaptability under computational constraints, we propose squeeze and excitation Conformer (SEConformer), a lightweight variant that systematically reconstructs both the convolutional branch and the FFN structure.

In the feed-forward module, a conventional FFN adopts an expand-reduce strategy, where the input dimension is first expanded by a factor of \(N\) and then projected back to its original size. While this increases representational capacity, it also incurs high computational cost due to large matrix multiplications. On closer inspection, we observe that the tail FFN in each Conformer block is usually followed by another FFN at the beginning of the next block. This configuration leads to repeated transformations of high-level features and causes unnecessary redundancy.
SEConformer addresses this issue by removing the tail FFN from the first block and redesigning the following FFN as a squeeze and excitation module. This module uses the Swish activation function to apply non-linear compression along the channel dimension, which helps the network focus on more relevant features and improves the efficiency of information processing, as shown in Fig.~\ref{fig:conformer}. This modification greatly reduces the number of parameters and the computational load, while still maintaining strong feature representation ability.

\begin{figure}[!t]
    \centering
    \includegraphics[width=\linewidth]{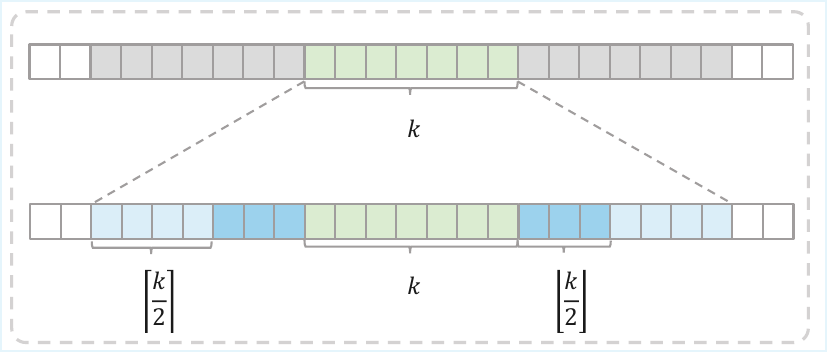}
    \caption{Schematic diagram of changes in the receptive field under the same convolutional kernel.}
    \label{fig:shiftconv_a}
\end{figure}

\begin{figure}[!t]
    \centering
    \includegraphics[width=\linewidth]{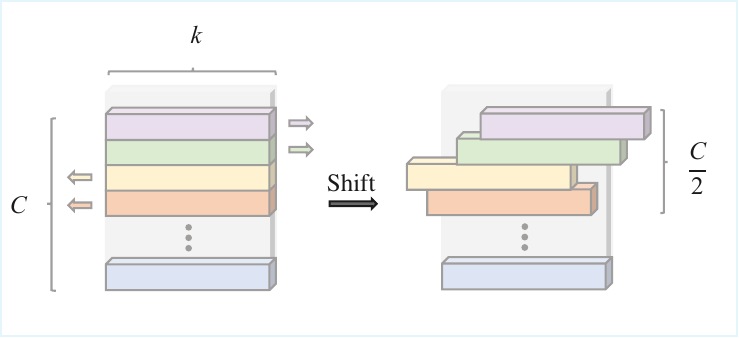}
    \caption{Schematic diagram of channel variation in time-shift convolution.}
    \label{fig:shiftconv_b}
\end{figure}

For the convolutional module, SEConformer incorporates a time-shift convolution to strengthen temporal modeling. Increasing kernel size is a straightforward way to capture long-range dependencies, but it often causes optimization and hardware inefficiencies. Inspired by shift mechanisms in the image domain~\cite{chengSkeletonBasedActionRecognition2020,chenAllYouNeed2019,liShiftwiseConvSmallConvolutional2025}, we use simple shift operations to enlarge the receptive field without introducing extra parameters.
The time-shift convolution works by shifting the input sequence forward and backward in time, then concatenating the shifted signals with the original features along the channel dimension. In this way, the model incorporates information across frames while keeping the kernel size unchanged. Although small kernels are still used, the receptive field is effectively enlarged, and the convolution can respond to temporal variations over a longer range, as illustrated in Fig.~\ref{fig:shiftconv_a}.
Standard convolutions usually rely on large kernels to broaden the receptive field. In contrast, time-shift convolution reaches a similar effect by combining small kernels with structured channel-wise shifts. This design improves the ability of model to capture long-term context while keeping the parameter count nearly unchanged.
For a time series with $C$ channels, only $C/2$ channels are selected for the shift operation to ensure that the primary receptive field of the convolution kernel remains centered on the current time step. These $C/2$ channels are further divided equally into four parts, to which forward and backward temporal shifts of varying lengths are applied, as depicted in Fig.~\ref{fig:shiftconv_b}. Specifically, the shift lengths are determined based on the convolution kernel size $k$ and defined as:\(+k, -k, +\lfloor k/2 \rfloor, -\lfloor k/2 \rfloor\). This design allows the model to collect contextual information at different time scales.

Overall, the time-shift convolution enhances temporal modeling through simple and efficient tensor operations. Unlike conventional approaches that expand the receptive field by enlarging kernels, it achieves comparable effects with minimal computational cost and no additional parameters.

\subsubsection{State space model optimization}
In moving sound source localization, the DOA changes continuously over time, resulting in features with strong temporal variation. The Mamba architecture has shown good potential in modeling long sequences by combining state space models with a selective scanning mechanism, which helps reduce the computational cost often seen in attention-based methods. However, the standard Mamba framework processes input data only in the forward direction, limiting its ability to capture full temporal dependencies. This drawback is especially noticeable in dynamic DOA estimation, where accurate localization requires context from both past and future frames. In addition, the selective recurrence design of Mamba tends to prioritize recent information, making it less effective at retaining long-range historical context, which further restricts its performance in tasks where long-term dependencies play a critical role.

\begin{figure}[!t]
    \centering
    \includegraphics[width=\linewidth]{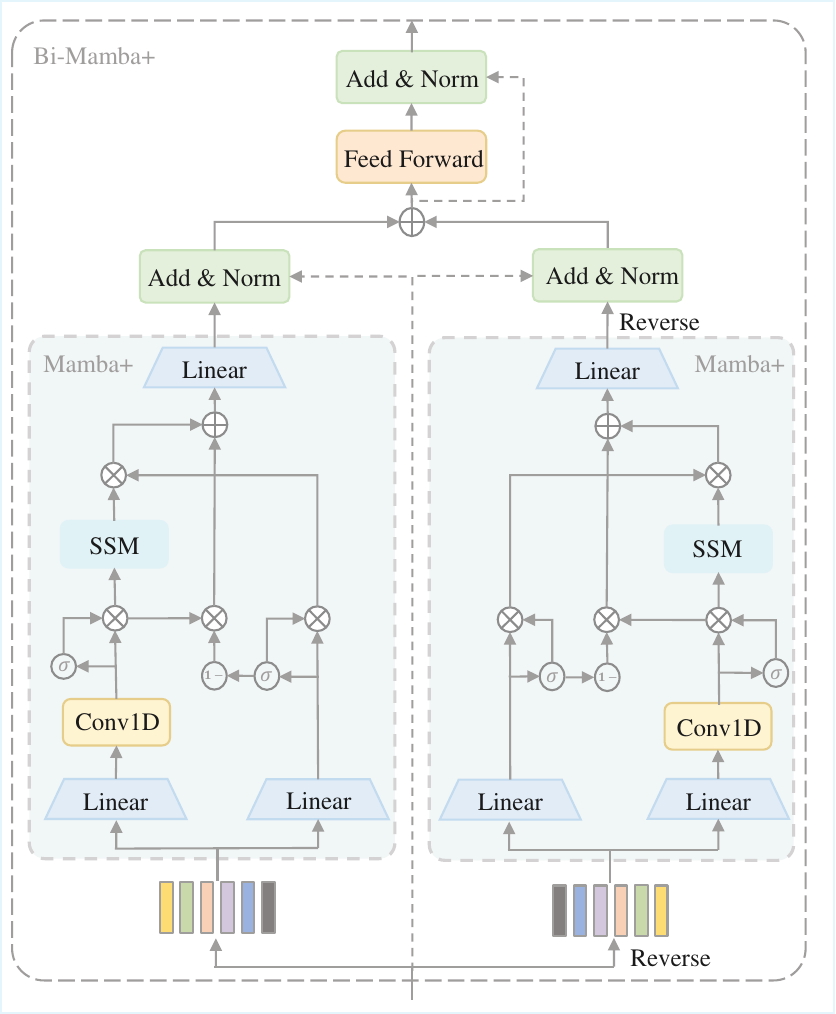}
    \caption{The illustrations of Bi-Mamba+.}
    \label{fig:mamba}
\end{figure}

To address this limitation, we introduce a bidirectional Mamba+ \cite{liangBiMambaBidirectionalMamba2024}architecture that strengthens contextual modeling by incorporating information from both past and future time steps. Unlike the standard Mamba, Mamba+ introduces a learnable forget gate within each branch to enable selective integration of new inputs with historical features, allowing the model to capture complementary temporal dependencies more effectively. The structure has two parallel branches: one processes the input sequence in the forward direction, and the other processes a reversed sequence in the backward direction. While both branches share the same overall design, they maintain separate state transitions so that temporal cues can be learned in both directions. The outputs from the forward and backward branches are then combined through a fusion module, yielding a unified representation that reflects richer spatiotemporal patterns across the entire sequence. As shown in Fig.~\ref{fig:mamba}, the Bi-Mamba+ follows a parallel dual-branch structure that is both efficient and easy to integrate into existing sound localization frameworks.

\begin{algorithm}
\caption{The process of Mamba+ Block}
\label{algorithm1}
\textbf{Input:} $\bm{X} : (B, L, D)$ \\
\textbf{Output:} $\bm{Y} : (B, L, D)$
\begin{algorithmic}[1]
\State $\bm{x}, \bm{z} : (B, L, ED) \gets \text{Linear}_X(\bm{X}), \text{Linear}_Z(\bm{X}) $
\State $\bm{x}' : (B, L, ED) \gets \text{SiLU}(\text{Conv1D}(\bm{x}))$ 
\State $\bm{A} : (D, N) \gets \text{Parameter}$ 
\State $\bm{B}, \bm{C} : (B, L, N) \gets \text{Linear}_B(\bm{x}'), \text{Linear}_C(\bm{x}')$
\State $\bm{\Delta} : (B, L, D) \gets \text{Softplus}\left(\text{Linear}_\Delta(x') + \text{Parameter}_\Delta\right)$
\State $\bar{\bm{A}}, \bar{\bm{B}} : (B, L, D, N) \gets \text{discretize}(\bm{\Delta}, \bm{A}, \bm{B})$ 
\State $\bm{y} : (B, L, ED) \gets \text{SSM}(\bar{\bm{A}}, \bar{\bm{B}}, \bm{C})(\bm{x}')$
\State $\bm{y}' : (B, L, ED) \gets \bm{y} \otimes \text{SiLU}(\bm{z}) + x' \otimes (1 - \sigma(z))$ 
\State $\bm{Y} : (B, L, D) \gets \text{Linear}_D(\bm{y}')$
\State \Return $\bm{Y}$
\end{algorithmic}
\end{algorithm}

Mamba+ represents all recurrent processes with hidden states through two sets of equations, as described in Algorithm~\ref{algorithm1}.
In continuous-time state space models, the system’s behavior in response to an input signal \( x(t) \in \mathbb{R} \) is described by the evolution of a hidden state \( h(t) \in \mathbb{R}^N \) over time. The dynamics can be formulated as:
\begin{equation}
\begin{split}
h'(t) &= A x(t) + B h(t), \\
y(t) &= C h(t),
\end{split}
\label{eq:mamba_ssm}
\end{equation}
where \( A \in \mathbb{R}^{N \times N} \) determines how the input affects the hidden state, \( B \in \mathbb{R}^{N \times 1} \) regulates the internal state transitions, and \( C \in \mathbb{R}^{1 \times N} \) maps the hidden state to the system output \( y(t) \in \mathbb{R} \).

Since digital systems process discrete-time signals,  Eq.~\eqref{eq:mamba_ssm} is typically discretized using the Zero-Order Hold~\cite{pechlivanidouZeroorderHoldDiscretization2022} method. For a fixed time interval $\Delta$, the discretized state-space equations become:
\begin{equation}
\begin{split}
\bar{A} &= \exp(\Delta A), \\
\bar{B} &= \left( \Delta A \right)^{-1} \left( \exp(\Delta A) - I \right) \Delta B.
\end{split}
\label{eq:zoh_discretization}
\end{equation}

Finally, the formula of discretized SSM can then be written as:
\begin{equation}
\begin{split}
h_t &= \bar{A} h_{t-1} + \bar{B} x_t, \\
y_t &= C h_t.
\end{split}
\label{eq:discrete_ssm}
\end{equation}
In the Bi-Mamba+ setting, these equations are applied independently in both forward and backward branches, and the resulting outputs are combined to form a richer representation.

\subsubsection{State-space and self-attention collaborative network}

\begin{figure}[!t]
    \centering
\includegraphics[width=0.6\linewidth]{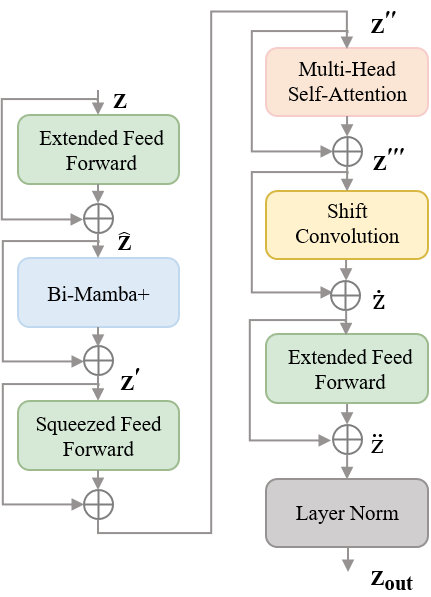}
    \caption{The illustration of Stateformer.}
    \label{fig:stateformer}
\end{figure}
To improve the global sequence modeling ability while preserving computational efficiency, we propose Stateformer, a new architecture built on the lightweight SEConformer framework and incorporating a state-space modeling module. As illustrated in Fig.~\ref{fig:stateformer}, the design combines the strengths of self-attention and state-space mechanisms. By replacing certain components and stacking multiple layers, Stateformer increases the depth of temporal modeling without significantly increasing the model size or computation cost.
Self-attention has been widely used in speech and audio tasks due to its ability to capture global context. However, its quadratic complexity with respect to sequence length makes it less suitable for long sequences in real-time or resource-limited scenarios. In contrast, Mamba, a model based on SSM, achieves linear time complexity, allowing efficient modeling of long-range dependencies and deeper networks with lower computational overhead. Nevertheless, Mamba is less effective in capturing fine local variations compared to self-attention.

To address these limitations, Stateformer combines both methods in a single architecture. The network keeps the overall structure of SEConformer but replaces the multi-head self-attention and convolutional modules in the first layer with a Bi-Mamba+ block. This modification improves the model’s ability to capture sequence-level patterns from the input layer. In addition, the structural bias introduced by the state space model helps the network to better capture long-term dependencies, which are often difficult for self-attention alone.

The specific process can be described by the following formulas:
\begin{equation}
\hat{\mathbf{Z}} = \mathbf{Z}+\frac{1}{2}\text{FFN}_\text{E}(\mathbf{Z}),
\end{equation}
\begin{equation}
\mathbf{Z}^\prime = \text{DyTanh}(\hat{\mathbf{Z}}+\text{Bi-Mamba}^+(\hat{\mathbf{Z}})),
\end{equation}
\begin{equation}
\mathbf{Z}^{\prime\prime} = \mathbf{Z}^\prime+\frac{1}{2}\text{FFN}_\text{S}(\mathbf{Z}^\prime),
\end{equation}
\begin{equation}
\mathbf{Z}^{\prime\prime\prime} = \mathbf{Z}^{\prime\prime}+\text{MHSA}(\mathbf{Z}^{\prime\prime}),
\end{equation}
\begin{equation}
\dot{\mathbf{Z}} = \mathbf{Z}^{\prime\prime\prime}+\text{ShiftConv}(\mathbf{Z}^{\prime\prime\prime}),
\end{equation}
\begin{equation}
\ddot{\mathbf{Z}} = \dot{\mathbf{Z}}+\frac{1}{2}\text{FFN}_\text{E}(\dot{\mathbf{Z}}),
\end{equation}
\begin{equation}
\mathbf{Z}_\text{out} = \text{LN}(\ddot{\mathbf{Z}}).
\end{equation}
Among them, $\text{FFN}_\text{E}(\cdot)$ and $\text{FFN}_\text{S}(\cdot)$ denote the expansion and compression feed-forward layers, respectively. $\text{Bi-Marma}^{+}(\cdot)$ refers to the bidirectional Mamba+ module, which models temporal dependencies in both directions. $\text{DyTanh}(\cdot)$ stands for the dynamic Tanh activation function, and $\text{ShiftConv}(\cdot)$ indicates the shift-based convolutional layer designed to enlarge the temporal receptive field. $\mathbf{Z}$ represents the input time series, while $\mathbf{Z}_\text{out}$ denotes the corresponding output sequence.

\section{Experiments and Discussions} 
\label{sec:Experiments and Discussions}
\subsection{Datasets}

The simulated dataset is created by convolving RIRs with clean speech source signals. Pure speech signals are randomly selected from the Librispeech development corpus~\cite{panayotovLibrispeechASRCorpus2015} for VAD processing. In addition to white noise, diffuse noise and pink noise conditions are also considered. The dataset generation parameters, detailed in Table~\ref{tab:simulate datasets}, are randomly sampled from uniform distributions within specified intervals. These parameters include signal-to-noise ratio (SNR), reverberation time (RT60), room dimensions, and azimuth angles. The RIRs of moving sources are generated using the gpuRIR toolbox~\cite{diaz-guerraGpuRIRPythonLibrary2021}, chosen for its computational efficiency and advanced acoustic modeling capabilities. Two microphones are placed with an inter-microphone distance of 8 cm, and their positions are randomly determined within the room, constrained to lie on the same horizontal plane as the sound source. The final dataset consists of 20,480 training samples, 2,048 validation samples, and 1,024 testing samples.

For the real-world dataset, we use the LOCATA corpus from the IEEE-AASP Challenge on Sound Source Localization and Tracking~\cite{eversLOCATAChallengeAcoustic2020}. The recordings were made in a real room with dimensions of 7.1~m~\(\times\)~9.8~m~\(\times\)~3~m and a RT60 of 0.55~s. The LOCATA dataset provides an objective benchmark for state-of-the-art algorithms in sound source localization and tracking, comprising recordings of various real-world scenarios with both single and multiple sources, together with ground-truth information on source and sensor positions. For evaluation, we select data from Tasks 3 and 5, as well as Tasks 4 and 6, as publicly available real-world test sets.

\begin{table}[!t]
\centering
\caption{Parameters of Simulated Data.}
\label{tab:simulate datasets}
\begin{tabular}{ccc}
\toprule
Parameter & Value  & Unit \\
\addlinespace[0.3em]
\hline
\addlinespace[0.3em]
SNR & -5 -- 15 & dB \\
RT60 & 0.2 -- 1.3 & s \\
Room Size & 4$\times$5$\times$3 -- 10$\times$8$\times$6 & m$^3$ \\
Azimuth & 0 -- 180 & \textdegree  \\
\bottomrule
\end{tabular}
\end{table}

\subsection{Evaluation Metrics}
For quantitative evaluation, we use two widely used performance measures.
The average of the absolute error (MAE) is used to reflect the error between the predicted value and the ground truth:
\begin{equation}
\label{metric_mae}
\text{MAE}(^{\circ}) = \frac{1}{K} \sum_{k=1}^{K} \left( \frac{1}{S_k} \sum_{s=1}^{S_k} \left| \hat{\theta}_k^s - \theta_k^s \right| \right),
\end{equation}
where \(K\) represents the total number of evaluation samples. For each sample \(k\) containing \(S_k\) speakers, \(\hat{\theta}_k^s\) is the estimated angle for the \(s\)-th speaker, and \(\theta_k^s\) is its corresponding ground-truth angle.

The Accuracy (Acc) is calculated as the percentage of correctly localized samples:
\begin{equation}
\text{Acc} (\%) = \frac{K_c}{K_s} \times 100,
\end{equation}
where \( K_s \) represents the total evaluated samples and \( K_c \) counts accurately localized samples. A sample is considered correctly localized only if the absolute error for every speaker within it (\(\vert \hat{\theta}_k^s - \theta_k^s \vert\)) is less than or equal to an angular threshold \(\lambda_\theta\). The \( \lambda_{\theta} \) is set as 5°, 10°, and 15° in our experiments.

\subsection{Training Setup and Baseline Methods}
In our experiments, the proposed method processes audio in the STFT domain using a 32-ms Hanning window and a 16-ms overlap. To extract 256-dimensional complex spectral features, a 512-point discrete Fourier transform is applied, with the sampling rate fixed at 16 kHz.
For model optimization, the model parameters are optimized using the Adam algorithm, starting with a learning rate of 0.001. If the validation loss stagnates, the learning rate is reduced by 20\%, and training continues for up to 80 epochs.
The model is trained to output the Cartesian coordinates for each source present in a given time frame. The fundamental training goal is to minimize the mean squared error (MSE) between these predicted coordinates and the actual ground-truth locations.
However, a significant challenge in multi-source localization is permutation ambiguity, where the model might correctly predict the locations of all sources but in an arbitrary or incorrect order. To resolve this, we incorporate permutation invariant training (PIT)~\cite{yuPermutationInvariantTraining2017}. 

To ensure fair comparisons and eliminate the influence of hardware variability, all experiments are conducted on the same machine equipped with a single NVIDIA GeForce RTX 4090 GPU. The detailed specifications of the experimental platform are listed in Table~\ref{tab:experiment_platform}. To verify the effectiveness of the proposed method, we selected five algorithms for comparative analysis with the proposed method: CRNN-R~\cite{cooremanCRNNbasedMultiDOAEstimator2023a}, ResNet-Conformer (RC)~\cite{dongExperimentalStudyJoint2025}, ConBiMamba~\cite{houConMambaConvolutionAugmentedMamba2024}, TF-Mamba~\cite{xiaoTFMambaTimeFrequencyNetwork2025} and IPDnet~\cite{wangIPDnetUniversalDirectPath2024a}. Among these, CRNN-R serves as the baseline algorithm. To ensure consistency, all compared models are retrained on the same simulated dataset used in our work. All these algorithms, including the proposed method, perform offline localization. For RC and ConBiMamba, we adapt their targets to multi-source DOAs in this experiment. To ensure consistency, all compared models are retrained on the same simulated dataset used in our work.

\begin{table}[!t]
\centering
\caption{Experimental Setup and System Configuration}
\begin{tabular}{ll}
\toprule
\textbf{Item} & \textbf{Details} \\
\midrule
Operating System  & Ubuntu 22.04 \\
Processor         & Intel Core i7-13700KF (13th Gen) \\
RAM               & 64 GB \\
GPU               & NVIDIA RTX 4090 (24 GB VRAM) \\
CUDA Toolkit      & v. 11.8 \\
Python            & v. 3.10.13 \\
PyTorch           & v. 2.3.1 \\
\bottomrule
\end{tabular}
\label{tab:experiment_platform}
\end{table}

\begin{enumerate}[label=\arabic*)]

    \item \textbf{CRNN-R}\cite{cooremanCRNNbasedMultiDOAEstimator2023a} uses a CRNN architecture for multi-source DOA regression. The original model only uses the phase spectrogram as input and we consider incorporating magnitude information.
    
    \item \textbf{RC}~\cite{dongExperimentalStudyJoint2025} won first place in the DCASE2024 challenge. It integrates ResNet blocks with the Conformer. In our experiments, we use microphone signals as its input features.
    
    \item \textbf{ConBiMamba}~\cite{houConMambaConvolutionAugmentedMamba2024} proposes to replace the multi-head self-attention of Conformer with an external bidirectional Mamba layer, enabling linear-time sequence modeling while retaining global receptive-field. 
   
    \item \textbf{IPDnet}\cite{wangIPDnetUniversalDirectPath2024a} introduces a full-band and narrow-band fused LSTM architecture to estimate the direct-path IPD (DP-IPD) information from microphone array signals, thereby enabling robust multi-source SSL.

    \item \textbf{TF-Mamba}~\cite{xiaoTFMambaTimeFrequencyNetwork2025} is a 2-microphone SSL model for single-source DOA estimation. It uses a bidirectional Mamba network to process temporal and frequency sequences jointly. The input is the real and imaginary parts of dual-channel STFT coefficients. 
\end{enumerate}

\subsection{Experiment on Localization Performance}

\subsubsection{Comparison with other methods}

\begin{table}[!t]
\centering
\small
\captionsetup{width=\linewidth}
\caption{Experimental Performance of FA-Stateformer and Comparative Methods Under Moving Speaker Conditions in the Simulated Dataset. Statistical Significance Is Indicated With * ($p<0.05$), ** ($p<0.01$), and *** ($p<0.001$), Compared to the Proposed FA-Stateformer Model.}
\label{tab:simulate}
\setlength{\tabcolsep}{3pt}
\resizebox{\linewidth}{!}{
\begin{tabular}{@{}ccccc@{}}
\toprule
Methods & Acc5 (\%)$\uparrow$ & Acc10 (\%)$\uparrow$ & Acc15 (\%)$\uparrow$ & MAE (\textdegree)$\downarrow$ \\
\midrule
CRNN-R & 52.5*** & 75.0*** & 84.4*** & 3.7*** \\
RC & 62.2*** & 84.6*** & 91.4** & 3.3** \\
ConBiMamba & 62.1*** & 83.5*** & 90.8** & 3.3** \\
IPDnet & 69.0* & 86.8 & 91.6** & 2.8 \\
TF-Mamba & 67.5** & 86.3* & 92.0* & 3.2* \\
FA-Stateformer & \textbf{71.8} & \textbf{88.1} & \textbf{93.7} & \textbf{2.7} \\
\bottomrule
\end{tabular}
}
\end{table}

Table~\ref{tab:simulate} presents the Acc under different thresholds and MAEs of each method in the moving two-speaker setting. The simulation results indicate that the proposed algorithm achieves the best performance among all compared methods.
Compared to CRNN-R methods, RC replaces RNN with Conformer to improve overall performance. 
RC is based on ResNet blocks and Conformer. The architecture of ResNet blocks is detailed as follows: four ResNet block progressively increase the number of channels from 4 to 32, 64, 128.
ConMamba replaces the multi-head self-attention mechanism of the Conformer with Mamba layers and adds convolutional layers to capture both local and global features. ConMamba achieves comparable performance to the traditional Conformer on short speech segments while effectively addressing computational complexity and positional awareness issues, as discussed in detail in Section~\ref{sec:Complexity}.
IPDnet utilizes a full-band and narrowband fusion network coupled with a multi-track DP-IPD learning objective to achieve excellent sound source localization performance. TF-Mamba adopts a similar concept of full-band and narrowband fusion, applying Mamba to both the time and frequency domains to build a dual-dimension approach.

The proposed FA-Stateformer achieves state-of-the-art performance among the compared methods. Compared with ConBiMamba, it integrates state space modeling with self-attention, which enables more effective exchange of information across both time and frequency. In contrast to CRNN-R and RC, FA-Stateformer reduces the risk of feature confusion and preserves cues that are important for localization. At the same time, it maintains a relatively small number of parameters, leading to more accurate DOA estimation.
To further confirm the reliability of the observed improvements, paired $t$-tests were conducted between FA-Stateformer and each comparative model over multiple experimental runs. The statistical results demonstrate that the performance gains of FA-Stateformer are statistically significant, with $p$-values less than 0.05, 0.01, and 0.001 in Table~\ref{tab:simulate}. These findings verify that the proposed model achieves superior localization accuracy and efficiency under moving-speaker conditions in the simulated dataset.

\begin{figure}[!t]
    \centering
    \includegraphics[width=0.9\linewidth]{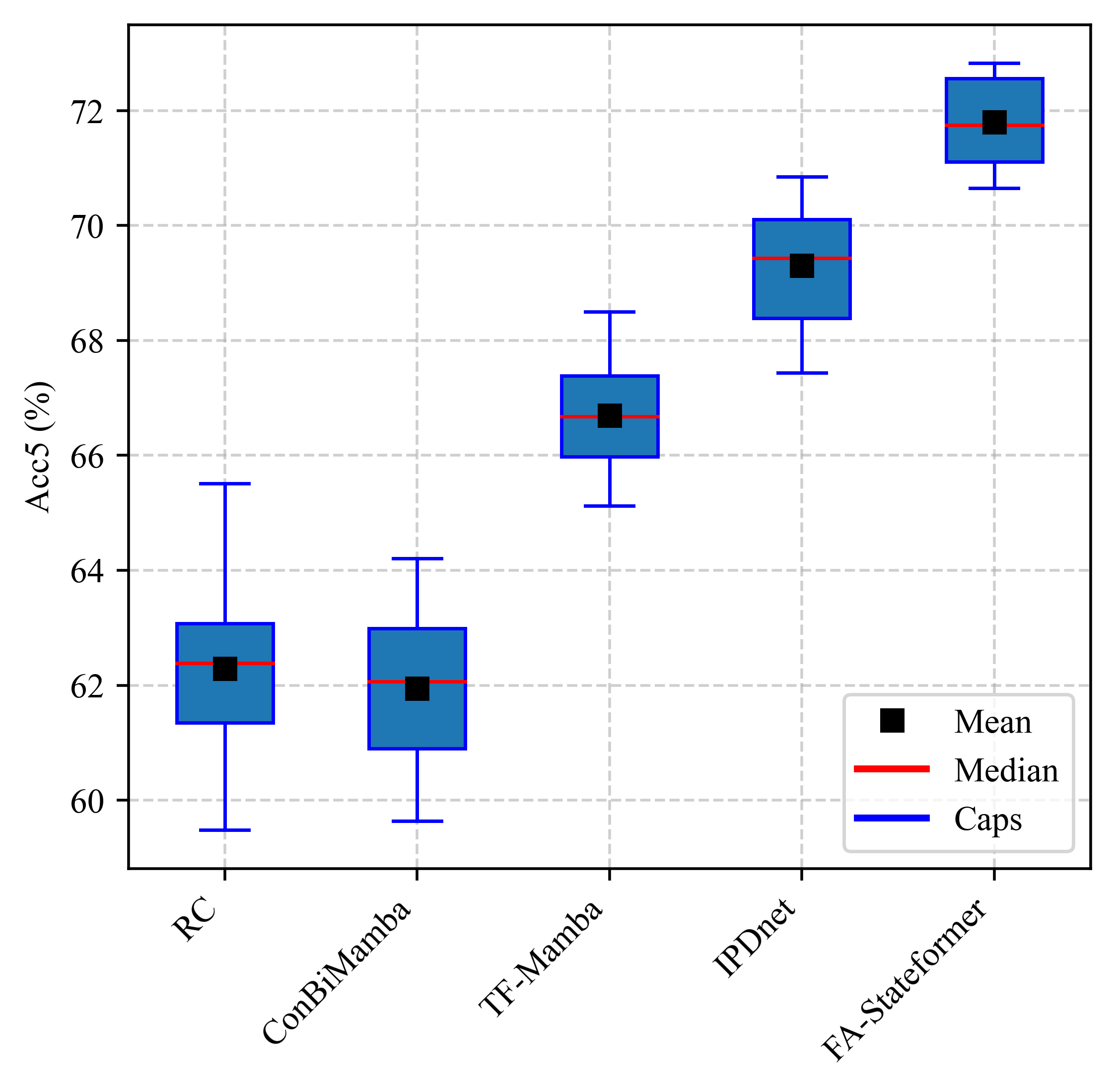}
    \caption{Acc5 comparison across methods based on 5$\times$2 cross-validation.}
    \label{fig:box}
\end{figure}

To further analyze the robustness of different methods on simulated datasets, we analyzed the accuracy of each method using 5$\times$2 cross-validation and visualized the results with boxplots. As shown in Fig.~\ref{fig:box}, these boxplots display key statistical metrics such as the mean, median, standard deviation, and variability of the results. The figure shows that the traditional methods RC exhibit significant performance fluctuations, with the overall distribution of Acc5 being low. TF-Mamba show some improvement in the median but still exhibits significant variability. Notably, FA-Stateformer achieves the highest performance and stability among all algorithms, with a significantly narrower box range and superior mean and median performance compared to the comparison methods, demonstrating the stability of the proposed method in complex moving speaker scenarios.

\subsubsection{Evaluation on different reverberant-noisy experiments}
\begin{table*}[!t]
  \centering
  \caption{Performance Comparison of Different Models on Simulated Datasets Under Varying SNR and RT60 Conditions}
  \label{tab:snr_rt60}
  \setlength{\tabcolsep}{2pt}
  \resizebox{\linewidth}{!}{
  \begin{tabular}{llcccccccccccc}
    \toprule
    \textbf{Methods} & & \multicolumn{2}{c}{Ours} & \multicolumn{2}{c}{IPDnet} & \multicolumn{2}{c}{TF-Mamba} & \multicolumn{2}{c}{ConBiMamba} & \multicolumn{2}{c}{RC} & \multicolumn{2}{c}{CRNN} \\
    \cmidrule(lr){3-4} \cmidrule(lr){5-6} \cmidrule(lr){7-8} \cmidrule(lr){9-10} \cmidrule(lr){11-12} \cmidrule(lr){13-14}
    \textbf{Metric} & & Acc5(\%)$\uparrow$ & MAE(°)$\downarrow$ & Acc5(\%)$\uparrow$ & MAE(°)$\downarrow$ & Acc5(\%)$\uparrow$ & MAE(°)$\downarrow$ & Acc5(\%)$\uparrow$ & MAE(°)$\downarrow$ & Acc5(\%)$\uparrow$ & MAE(°)$\downarrow$ & Acc5(\%)$\uparrow$ & MAE(°)$\downarrow$ \\
    \midrule
    \multirow{7}{*}{SNR Levels (dB)} 
    & -10 & 56.2 & 2.2 & 51.8 & 2.1 & 50.4 & 2.2 & 47.3 & 2.3 & 48.7 & 2.3 & 40.3 & 2.3 \\
    & -5 & 63.7 & 2.1 & 59.5 & 2.1 & 56.9 & 2.2 & 54.5 & 2.2 & 55.7 & 2.3 & 45.7 & 2.3 \\
    & 0 & 68.9 & 2.1 & 65.4 & 2.1 & 63.7 & 2.1 & 59.6 & 2.2 & 60.3 & 2.2 & 50.2 & 2.3 \\
    & 5 & 72.2 & 2.0 & 69.4 & 2.0 & 67.2 & 2.1 & 62.5 & 2.2 & 63.1 & 2.2 & 53.3 & 2.3 \\
    & 10 & 74.5 & 2.0 & 72.3 & 1.9 & 70.1 & 2.0 & 64.5 & 2.2 & 64.6 & 2.2 & 55.5 & 2.3 \\
    & 15 & 75.7 & 1.9 & 74.1 & 1.8 & 72.6 & 2.0 & 65.7 & 2.1 & 65.5 & 2.2 & 56.9 & 2.3 \\
    & 20 & 76.5 & 1.9 & 75.9 & 1.8 & 73.7 & 1.9 & 66.3 & 2.1 & 65.9 & 2.2 & 57.6 & 2.1 \\
    \cmidrule(lr){2-14}
    & Avg & \textbf{69.7} & \textbf{2.0} & 66.9 & \textbf{2.0} & 64.9 & 2.2 & 60.1 & 2.2 & 60.5 & 2.2 & 51.4 & 2.3 \\
    \midrule
    \multirow{7}{*}{RT60 (s)} 
    & 0.2 & 76.2 & 1.7 & 74.6 & 1.7 & 70.5 & 2.0 & 64.9 & 2.1 & 65.8 & 2.2 & 56.8 & 2.2 \\
    & 0.3 & 73.9 & 1.7 & 72.7 & 1.7 & 69.2 & 2.0 & 63.7 & 2.1 & 64.1 & 2.2 & 55.7 & 2.2 \\
    & 0.4 & 73.5 & 1.7 & 71.2 & 1.8 & 68.5 & 2.1 & 63.2 & 2.2 & 64.0 & 2.2 & 54.9 & 2.3 \\
    & 0.5 & 72.6 & 1.8 & 69.3 & 1.8 & 67.4 & 2.1 & 62.1 & 2.2 & 63.0 & 2.2 & 53.1 & 2.3 \\
    & 0.6 & 71.2 & 1.8 & 67.6 & 1.9 & 66.2 & 2.1 & 60.7 & 2.2 & 61.5 & 2.2 & 51.6 & 2.3 \\
    & 0.7 & 70.1 & 1.9 & 66.2 & 1.9 & 65.3 & 2.2 & 59.5 & 2.2 & 60.4 & 2.3 & 50.4 & 2.3 \\
    & 0.8 & 68.7 & 1.9 & 64.4 & 2.0 & 64.5 & 2.2 & 58.6 & 2.3 & 58.9 & 2.3 & 48.9 & 2.3 \\
    \cmidrule(lr){2-14}
    & Avg & \textbf{72.3} & \textbf{1.8} & 69.3 & \textbf{1.8} & 67.3 & 2.1 & 61.8 & 2.2 & 62.5 & 2.2 & 53.1 & 2.3 \\
    \bottomrule
  \end{tabular}
  }
\end{table*}

As illustrated in Table~\ref{tab:snr_rt60}, localization performance under threshold 5° is evaluated across a range of RT60 values (0.2 -- 0.8 s) and SNR levels (-10 -- 20 dB). 
The proposed FA-Stateformer consistently achieves the best overall accuracy and the lowest MAE across all conditions.
Although FA-Stateformer surpasses IPDnet in overall accuracy, the two methods achieve similar MAE values. The DP-IPD learning objective used in IPDnet directly enhances sensitivity to phase differences and stabilizes angle estimation, which helps the network maintain low prediction error. However, this design lacks the broader context modeling capability offered by FA-Stateformer.

\subsubsection{Model performance in real-world environments}

In this subsection, we evaluate the proposed model using the LOCATA challenge database, which provides a benchmark for sound source localization under realistic acoustic conditions. The experiments focus on Tasks 3–6. Tasks 3 and 5 contain recordings with a single dynamic sound source, where the number of sources is known. Task 3 focuses on scenarios where the speaker moves and may also rotate the head and body, allowing the study of source direction changes under controlled conditions. Task 5 provides a fully dynamic setting in which both the source and the microphone array are moving, creating a more complex real-world scenario. Tasks 4 and 6 involve recordings with multiple sources, where the number of active sources is not known in advance. Task 6 is particularly challenging, as it includes multiple moving speakers recorded with a moving microphone array, resulting in highly dynamic acoustic scenes.

\begin{figure*}[!t]
    \centering
    \includegraphics[width=\linewidth]{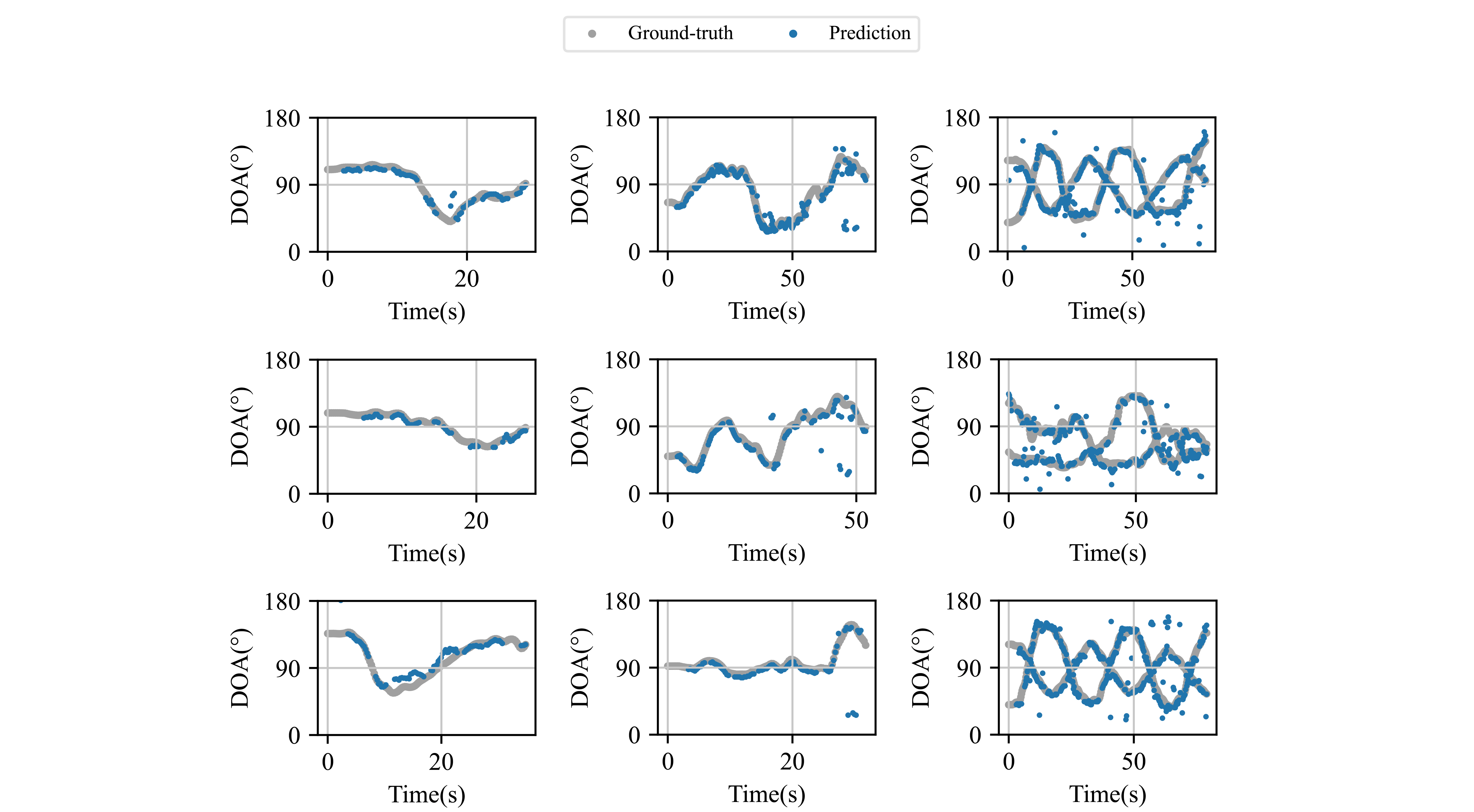}
    \caption{DOA estimation examples from LOCATA dataset.}
    \label{fig:locata}
\end{figure*}

Table~\ref{tab:locata} presents the Acc10, Acc15, and MAE results on LOCATA Tasks 3--6. Compared with other methods, the proposed approach achieves higher accuracy in both single-speaker and two-speaker scenarios. To supplement the numerical results, several localization examples obtained by the proposed model are shown in Fig.~\ref{fig:locata}, giving a more intuitive view of the prediction performance.  
It should be noted that the training uses a two-microphone planar array, which limits DOA estimation to a 180-degree azimuth range. However, the LOCATA dataset covers the full range of \([-180^{\circ},180^{\circ}]\). This limitation has a strong impact on Task~6, where both the sources and the array move with large azimuth changes. Consequently, all models, including the proposed one, show reduced accuracy on this task. Nevertheless, the proposed method still achieves better performance than the baseline methods.

\begin{table*}[!t]
\centering
\small               
\caption{Azimuth Localization Performance on the LOCATA Dataset}
\label{tab:locata}
\setlength{\tabcolsep}{2pt}
\resizebox{\linewidth}{!}{
\begin{tabular}{ccccccccccccc}
\toprule
\multirow{2}{*}{{Methods}} & \multicolumn{3}{c}{{Task3}} & \multicolumn{3}{c}{{Task4}} & \multicolumn{3}{c}{{Task5}} & \multicolumn{3}{c}{{Task6}} \\
\cmidrule(lr){2-4} \cmidrule(lr){5-7} \cmidrule(lr){8-10} \cmidrule(lr){11-13}
 & {Acc10(\%)$\uparrow$} & {Acc15(\%)$\uparrow$} & {MAE(°)$\downarrow$} & {Acc10(\%)$\uparrow$} & {Acc15(\%)$\uparrow$} & {MAE(°)$\downarrow$} & {Acc10(\%)$\uparrow$} & {Acc15(\%)$\uparrow$} & {MAE(°)$\downarrow$} & {Acc10(\%)$\uparrow$} & {Acc15(\%)$\uparrow$} & {MAE(°)$\downarrow$} \\
\midrule
CRNN-R & 87.5 & 94.9 & 3.6 & 66.5 & 81.0 & 4.4 & 62.4 & 70.0 & 3.9 & 34.7 & 45.3 & 4.5\\ 
RC & 94.4 & 96.5 & 3.4 & 68.2 & 82.6 & 4.3 & 76.4 & 79.0 & 3.7 & 55.6 & 62.1 & 3.8\\
ConBiMamba & 96.6 & 98.7 & 2.8 & 69.0 & 83.4 & 4.0 & 75.1 & 79.2 & 3.2 & 56.1 & 63.5 & 3.4 \\ 
IPDnet & 94.9 & 97.5 & 2.2 & 88.1 & 92.4 & 2.6 & 76.8 & 79.0 & \textbf{1.8} & 58.5 & 64.8 & \textbf{2.9} \\
TF-Mamba & 94.7 & 97.1 & 2.4 & 86.4 & 92.6 & 2.6 & 76.4 & 79.0 & 2.3 & 56.1 & 63.9 & 3.2 \\
FA-Stateformer & \textbf{95.7} & \textbf{99.0} & \textbf{2.1} & \textbf{92.9} & \textbf{94.1 }& \textbf{2.5} & \textbf{77.3} & \textbf{79.8} & 2.1 & \textbf{59.7} & \textbf{65.6} & \textbf{2.9} \\ 
\bottomrule
\end{tabular}
}
\end{table*}

\subsection{Ablation Study}
To further verify and analyze the effectiveness of the modules in the proposed architecture, ablation studies are conducted on the simulated data. We preserve identical hyperparameter values and consistent experimental configurations throughout the process. All results yield $p$-values from the $t$-test less than 0.05, demonstrating the statistical significance and effectiveness of our ablation studies. These findings validate the effectiveness of our method.

\subsubsection{The effect of the FA block}

\begin{table}[!t]
    \centering
\caption{Ablation Study on the FA Block}
\label{tab:RESULTS OF ABLATION TFA}
\setlength{\tabcolsep}{3pt}
\resizebox{\columnwidth}{!}{
    \begin{tabular}{@{}ccccc@{}}
        \toprule
        Methods & Params.(M) & Acc5 (\%)$\uparrow$ & Acc10 (\%)$\uparrow$ & MAE (°)$\downarrow$ \\
        \midrule
        Baseline  & \textbf{1.6} & 63.4 & 84.6 & 3.3 \\
        + TD  & 1.9 & 66.5 & 86.8 & 3.2 \\
        + FD  & 1.9 & 69.2 & 87.4 & 3.1 \\
        + FA block  & 2.2 & \textbf{71.8} & \textbf{88.1} & \textbf{2.7} \\
        \bottomrule
    \end{tabular}
    }
\end{table}

Table~\ref{tab:RESULTS OF ABLATION TFA} shows the DOA estimation results under two-source conditions on the simulated dataset.
The baseline model consists of a ResNet feature extractor followed by a Stateformer localization module, without any time or frequency dimension attention mechanisms.
Adding time-dimension (TD) attention improves performance compared with the baseline, particularly at smaller angular thresholds, which shows that modeling time information helps the network capture speech dynamics. Frequency-dimension (FD) attention brings even stronger gains, with clear improvements in both accuracy and MAE, indicating that spectral cues play a more critical role in localization. When both mechanisms are applied together, the model achieves the best results overall, with consistent accuracy gains and the lowest MAE. This demonstrates that combining time and frequency attentions enables the model to better exploit complementary cues for more reliable DOA estimation.

\subsubsection{Multiple configuration options for feedforward networks}

\begin{table}[!t]
    \centering
    \caption{Comparison of the Performance of Different FFN Settings}
    \setlength{\tabcolsep}{3pt}
    \resizebox{\columnwidth}{!}{
        \begin{tabular}{@{}ccccc@{}}
            \toprule
            Methods & Params.(M) & Acc5(\%)$\uparrow$ & Acc10(\%)$\uparrow$ & MAE(°)$\downarrow$ \\ \midrule
            FFN(F)      & 2.4 & 69.1 & 87.3 & 2.9 \\
            FFN(B)      & 2.4 & 70.2 & 87.6 & 2.8 \\
            FFN(FB)     & 1.9 & 67.4 & 85.6 & 3.1 \\
            SEConformer & \textbf{2.0} & \textbf{70.9} & \textbf{87.8} & \textbf{2.8} \\ 
            \bottomrule
        \end{tabular}
    }
    \label{tab:FFN}
\end{table}

We conducted ablation experiments on different configurations of feedforward networks to examine how design choices affect model performance. The results are reported in Tables~\ref{tab:FFN} . The RC network was used as the baseline, and three FFN compression strategies were evaluated: FFN(F), FFN(B), and FFN(FB). Specifically, FFN(F) compresses the forward FFN module in the Conformer, FFN(B) compresses the backward module, and FFN(FB) applies compression to both modules, replacing the conventional expansion scheme. In addition, SEConformer introduces lightweight modifications to the overall architecture, reducing model parameters and computational cost. 

\subsubsection{Performance comparison of different ShiftConv settings}
In the experimental exploration of the performance of the Conformer architecture, we also designed three ablation networks, namely Conv31, Conv21, and Conv15, and conducted comparative analysis by gradually reducing the convolution kernel size, as shown in Table~\ref{tab:ShiftConv}. 
\begin{table}[!t]
    \centering
    \caption{Comparison of the Performance of Different ShiftConv Settings}
    \begin{tabular}{@{}cccc@{}}
        \toprule
        Methods & Acc5(\%)$\uparrow$ & Acc10(\%)$\uparrow$ & MAE(°)$\downarrow$ \\ \midrule
        Conv31      & 71.2 & 88.1 & 2.7 \\
        Conv21      & 70.6 & 88.0 & 2.8  \\
        Conv15      & 70.4 & 87.7 & 2.8 \\
        ShiftConv31 & 71.0 & \textbf{88.2} & 2.7 \\
        ShiftConv15 & \textbf{71.8} & 88.1 & \textbf{2.7} \\ 
        \bottomrule
    \end{tabular}
    \label{tab:ShiftConv}
\end{table}
Experimental results show that as the convolution kernel size decreases, network performance shows a clear downward trend: for example, the Acc5 of Conv15 drops from 71.2\% for Conv31 to 70.4\%. This result demonstrates that traditional large convolution kernels have stronger feature extraction capabilities in sequence modeling. Simply reducing the convolution kernel size weakens the network's ability to capture temporal information, leading to performance degradation. To overcome this limitation, we further introduce a time-shifted convolution mechanism to construct two lightweight networks, ShiftConv21 and ShiftConv15. Through the ShiftConv operation, the receptive field is effectively expanded without increasing the number of parameters. Specifically, ShiftConv15 achieves a 2.0\% improvement over Conv15, outperforming not only the small-kernel network but also the large-kernel Conv31 model. 
These results demonstrate that the ShiftConv mechanism can effectively compensate for the reduced receptive field caused by smaller kernels, substantially enhancing the model’s capability for long sequence temporal modeling.

\subsection{Complexity Analysis}
\label{sec:Complexity}
\begin{table}[!t]
    \centering
    \caption{Comparison of Computational Complexity}
    \setlength{\tabcolsep}{3pt}
    \resizebox{\linewidth}{!}{
    \begin{tabular}{@{}ccccccc@{}}
        \toprule
        Methods & Params.(M) & FLOPs (G/s)$\downarrow$ & Times (ms)$\downarrow$ & Acc5 (\%)$\uparrow$ \\
        \midrule
        CRNN-R & \textbf{0.7} & \textbf{2.2} & 88.3 & 52.5 \\
        RC & 4.3 & 6.4 & 119.4 & 62.2 \\
        ConBiMamba & 2.5 & 4.7 & 9.6 & 62.1 \\
        IPDnet & 1.8 & 54.3 & 707.9 & 69.0 \\
        TF-Mamba & 2.0 & 43.8 & 125.2 & 67.5 \\
        FA-Stateformer & 2.2 & 4.7 & \textbf{9.5} & \textbf{71.8}  \\
        \bottomrule
    \end{tabular}
    }
    \label{tab:complexity}
\end{table}

To provide a comprehensive comparison of computational complexity, Table~\ref{tab:complexity} reports the number of parameters, floating point operations (FLOPs), inference time, and localization accuracy for CRNN-R, RC, ConBiMamba, TF-Mamba, IPDnet, and the proposed FA-Stateformer. All experiments are conducted on a system equipped with an Intel Core i7-13700KF CPU, 32 GB of memory, and an NVIDIA GeForce RTX 4090 GPU. FLOPs are measured with a batch size of 1.

From the results, FA-Stateformer shows clear advantages in both parameter count and FLOPs compared with the other networks. In particular, it achieves better accuracy with significantly lower computational cost than IPDnet, reducing FLOPs by nearly ten times while maintaining similar performance.
IPDnet and TF-Mamba use full-band and narrow-band networks to independently process frames and frequencies, respectively. This requires multiple network runs, resulting in higher overall complexity. However, TF-Mamba demonstrates its advantages over IPDnet in inference. When compared with the RC network, FA-Stateformer lowers FLOPs by 26.6\% and and further achieves an accuracy gain of 9.6\%. With respect to inference time, FA-Stateformer performs on par with ConBiMamba and is substantially faster than the commonly used RC network, with the inference time reduced by more than 90\%. These improvements are largely attributed to the use of the shift-convolution mechanism and compressed excitation patterns, which makes the SEConformer blocks more efficient and lightweight than standard Conformer blocks. 
This design not only reduces time and space complexity but also enhances the feature extraction capacity, allowing FA-Stateformer to achieve a better balance between accuracy and efficiency.

\section{Conclusion}
\label{sec:Conclusion}
In this paper, we proposed FA-Stateformer and evaluated its effectiveness for multi-speaker DOA estimation. FA-Stateformer combines a feature aggregation module with Stateformer blocks to efficiently model both temporal and frequency information. The feature aggregation module improves representation quality by emphasizing task-relevant features, while the Stateformer employs squeezed feed-forward layers and time-shift convolutions to achieve efficient sequence modeling. Compared with existing methods, FA-Stateformer shows clear advantages, achieving higher localization accuracy with fewer parameters and lower computational cost. For example, it improves accuracy by 9.6\% over RC while reducing FLOPs by 26.6\%, and maintains similar accuracy to IPDnet with only one-tenth of its computational load.  
Extensive experiments on both simulated and real-world datasets further confirm the effectiveness of FA-Stateformer. 

Although FA-Stateformer achieves a good balance between accuracy and efficiency, there are still open challenges. Future studies will focus on extending this framework to broader audio-related downstream tasks, such as sound event localization, separation, and speech enhancement. Another direction is to improve model robustness in diverse acoustic environments through domain adaptation and data augmentation. In addition, handling highly dynamic spatial scenes with moving sources and arrays remains a key challenge, and further advances in feature extraction and sequence modeling may help achieve stronger performance while keeping computational cost manageable.


\bibliographystyle{IEEEtran}
\bibliography{sci2} 


\newpage

 




\vfill

\end{document}